\documentclass[%
 reprint, superscriptaddress,
amsmath,amssymb,longbibliography,aps,prx,
article]{revtex4-2}
\pdfoutput=1
\usepackage{graphicx,multirow}%
\usepackage{amsmath}
\usepackage{gensymb}
\usepackage{bbm}
\usepackage[inkscapelatex=false]{svg}
\usepackage{dcolumn}
\usepackage{bm}
\usepackage{xcolor,colortbl}
\definecolor{blue}{rgb}{  0,  0,    1}
\definecolor{d4blue}{rgb}{  0,  0.4470,    0.7410}
\definecolor{d12orange}{rgb}{0.8500    0.3250    0.0980}
\definecolor{g8yellow}{rgb}{0.    0.6    0.298}
\definecolor{ppurple}{rgb}{0.4940    0.1840    0.5560}

\newcolumntype{k}{>{\columncolor{blue!20}}c}
\newcolumntype{r}{>{\columncolor{blue!10}}c}
\newcolumntype{d}{>{\columncolor{red!10}}c}
\usepackage{comment}

\newcommand{\down}{\downarrow}
\newcommand{\up}{\uparrow}

\usepackage{hyperref}

\begin{document}

\title{
Microscopic  Modeling of Interlayer Tunneling in Twisted Bilayer Cuprates}

\author{Siddhant Panda}
\affiliation{Department of Physics, University of Florida, Gainesville, Florida 32611, USA}
\author{Andreas Kreisel}
\affiliation{Niels Bohr Institute, University of Copenhagen, DK-2100 Copenhagen, Denmark}
\affiliation{Department of Physics and Astronomy, Uppsala University, Box 524, 751 20 Uppsala, Sweden}
\author{Laura Fanfarillo}
\affiliation{Istituto dei Sistemi Complessi (ISC-CNR), Via dei Taurini 19, I-00185 Rome, Italy}
\author{P. J. Hirschfeld}
\affiliation{Department of Physics, University of Florida, Gainesville, Florida 32611, USA}%

\date{\today}

\begin{abstract}
The field of twistronics, recently extended to twisted bilayer cuprate systems, has generated considerable interest due to the prediction of \mbox{$d+id'$}  states close to a $45\degree$ twist, as suggested by multiple experimental probes. Theoretical studies of these systems using a variety of models and many-body theoretical techniques have revealed multiple competing order parameters. The large variability in experimental results and theoretical predictions demands an accurate microscopic model for these systems, in particular a more realistic description of the interlayer tunneling. In this study, we fit the in-plane hopping and compute the interlayer tunneling amplitudes by calculating the orbital overlap between $d_{x^2-y^2}$ wave functions at each site of the twisted square-lattice model.
 We observe sign-changing tunneling amplitudes, a strong twist-angle dependence of the direct tunneling term, and off-vertical tunneling amplitudes that dominate over direct tunneling close to a $45\degree$ twist. We investigate the  band structure, Fermi surfaces, and spectral functions in comparison with those obtained from the phenomenological model and find a substantial redistribution of spectral weight that modifies the competition between different superconducting ground states. We determine the phase diagram as a function of twist angle, filling, and maximum tunneling strength, which is controlled by the $c$-axis distance between the monolayers, and observe a strong reduction in the region of stable time-reversal-symmetry-breaking (TRSB) \mbox{$d+id'$} states compared with the phenomenological model. Finally, we investigate the Josephson critical current across the phase diagram and find strongly suppressed critical currents close to a $45\degree$ twist, consistent with experimental observations.
\end{abstract}

\maketitle

\section{\label{sec:intro} Introduction}

The prediction of a time-reversal-symmetry-breaking (TRSB) chiral \mbox{$d+id'$} state in twisted bilayer cuprates \cite{Can2021,Kuboki1996,Kuboki1998} has generated considerable interest because of its potential applications in various quantum devices \cite{Patel2024,Confalone2025,Brosco2024}. The high $T_c$ of BSCCO  and its van der Waals interlayer interactions enabling atomically smooth cleaves make them an ideal platform for creating twist junctions. Tunneling experiments along the $c$ axis of BSCCO \cite{Li1999,Takano2002,Latyshev2004,Lee2021,FrankZhao,wang2023} have shown considerable variability in the measured Josephson critical current ($J_c$), with samples exhibiting either a persistent $J_c$ across all twist angles or a nearly vanishing $J_c$ close to $45\degree$ twist. Many of these twist junctions were fabricated and annealed at high temperatures \cite{Lee2021,Xikunxue}, a procedure that could potentially degrade their quality and lead to irreproducible results.
\par
More recently, cryogenic exfoliation techniques have enabled twist junctions to be fabricated at low temperatures, resulting in significantly reduced disorder~\cite{FrankZhao,MARTINI,Confalone2025Small,wang2023,zhu2025manipulatingfractionalshapirosteps,Patil2024}. However, these experiments have still exhibited some sample-to-sample variability, with some measurements reporting nearly constant $J_c$ as a function of twist angle \cite{wang2023,zhu2025manipulatingfractionalshapirosteps}.   To trace the origin of this variability, recent studies \cite{FrankZhao,MARTINI,Confalone2025Small} have tried to benchmark samples by measurements of their critical temperature, Josephson current density, and junction voltage. Junctions with the highest scores consistently display signatures compatible with TRSB \mbox{$d+id'$} states, including a strong twist-angle dependence of $J_c$, Shapiro steps, and Fraunhofer patterns. SQUID devices fabricated from these twisted interfaces have provided further evidence for \mbox{$d+id'$} order-parameter symmetry near a $45\degree$ twist angle \cite{Basu2026}.
\par
Theoretical studies of twisted bilayer cuprates have employed Ginzburg--Landau theory, continuum model descriptions, and lattice model calculations. Ginzburg--Landau theory and continuum models have the advantage of providing access to all twist angles but do not explicitly incorporate the microscopic details of the material. Both approaches have been central to the prediction of the \mbox{$d+id'$} state \cite{Can2021,Liu2023} and to subsequent studies of experimental observables, such as the field and angle dependence of the Josephson critical current \cite{Volkov2024,Volkov2025,PhysRevB.105.064501,PhysRevB.108.L100505}, the effects of incoherent tunneling \cite{Haenal2022}, quantum fluctuations \cite{shi2026quenchchiralsuperconductivityquantum}, and the robustness of the topological state \cite{li2026topologicalmulticomponentsuperconductivitysizable}. More recently, these approaches have been used to develop a general theory of twisted nodal superconductors, where square-lattice homobilayers with small twist angles  can result in fertile quantum simulators for extended Hubbard models \cite{Volkov2023,Luo2025,PixleyVolkov_review}.

\par
Lattice model descriptions are computationally expensive but provide greater control over the microscopic details of the system \cite{Can2021,panda2026,Senechaltwist2022,Senechaltwist2026,Fidrysiak2023,Song2022,PhysRevB.110.134514,PhysRevLett.127.157001}. Phase diagrams for twisted bilayer cuprates obtained from various lattice models have established the competition between multiple superconducting (SC) order parameters. In our previous study \cite{panda2026}, we found that the \mbox{$d+id'$} region at twist angles close to $45\degree$ lies in proximity to TRSB $d+is$ states, which are nontopological. We suggested that variations in the experimental fabrication process can alter the doping of the material, while the presence of disorder can enhance interlayer tunneling. Both effects could potentially explain why some twisted interfaces do not exhibit signatures of the topological \mbox{$d+id'$} state near a $45\degree$ twist.

\par
In our previous study \cite{panda2026}, we investigated the normal-state properties driving the different SC states found in our phase diagram, with the   Fermi surface spectral weight serving as a key probe for understanding the  symmetry of  SC ground states in different parameter regimes. The presence of substantial spectral weight in the antinodal region of a particular order-parameter symmetry lowers the condensation energy, thereby favoring that state. The sensitivity of the SC state to the normal-state properties suggests that the phase diagram can vary significantly between different models of these materials. Therefore, a microscopic model that incorporates the key properties and symmetries of BSCCO bilayers is essential.

\par
The intralayer normal-state properties of BSCCO, such as the nearest-neighbor and next-nearest-neighbor hopping parameters, are well established and have been extensively benchmarked. The interlayer tunneling terms, on the other hand, have often been simplified, particularly with regard to capturing the symmetries and orbital overlaps within the $\text{CuO}_2$ planes. Most studies have employed a phenomenological form of interlayer tunneling \cite{Can2021,panda2026,Senechaltwist2022,Fidrysiak2023} that decays exponentially with distance. Song \textit{et al.}~\cite{Song2022} studied twisted double-bilayer cuprates within a slave-boson mean-field treatment of the $t$-$J$ model, incorporating material-specific properties of BSCCO through different tunneling processes and their momentum dependence. They considered three tunneling processes: tunneling within the unit cell of each bilayer, direct tunneling between the bilayers, and tunneling mediated by the Cu $s$-orbital. They obtained a narrow region of TRSB but nontopological states around a $45\degree$ twist. These results further indicate that incorporating the microscopic details of these systems can substantially alter the phase diagram.
\par
The low-energy electronic structure of the cuprates can be obtained by downfolding the Cu--O multiorbital electronic structure onto an effective single-band description. The resulting low-energy Wannier orbital is centered on the Cu sites and has predominantly $d_{x^2-y^2}$ symmetry \cite{Andersen1995,Kreisel2015,Choubey2017}. Motivated by this microscopic orbital structure, we retain the real-space $d_{x^2-y^2}$ symmetry of the low-energy wave function by defining a $d_{x^2-y^2}$ wave function at each site of our square-lattice model. Modeling of interlayer tunneling has been previously done extensively in the context of twisted bilayers such as graphene and hexagonal boron nitride \cite{PhysRevB.93.235153,PhysRevB.85.195458}. We evaluate the orbital overlap between different sites in our moir\'e lattice for various twist angles and obtain a tunneling matrix with sign-changing tunneling amplitudes, which are absent in the phenomenological tunneling model. We also find that off-vertical tunneling amplitudes, between sites displaced by about one lattice constant, dominate over the direct tunneling term for twist angles close to $45\degree$. These differences lead to  band structures and Fermi surfaces distinct from those obtained using phenomenological tunneling. We investigate the phase diagram as a function of filling, twist angle, and interlayer tunneling strength, which is controlled by the $c$-axis distance between the monolayers, and obtain a substantially smaller region of TRSB \mbox{$d+id'$} states than that obtained with phenomenological tunneling. Finally, we calculate $J_c$ and demonstrate that the Josephson critical current vanishes as the twist angle approaches $45\degree$.

\section{Model}
\begin{figure}[tb]
\centering
\includegraphics[width=0.95\linewidth]{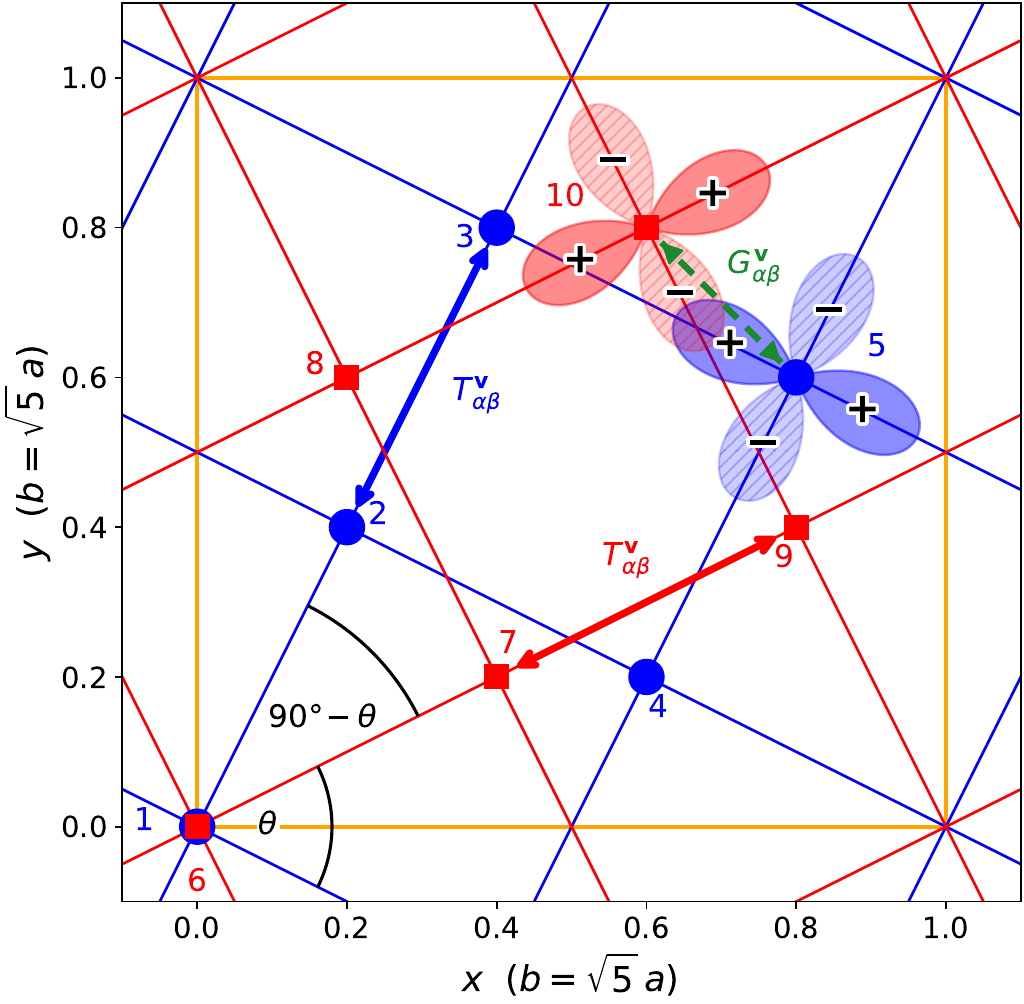}
\caption{Moir\'e unit cell for $(m,n)=(1,2)$, corresponding to a twist angle $\theta=53.13\degree\equiv36.87\degree$ due to the $C_4$ symmetry of the twisted square-lattice system. Each monolayer contains five lattice sites, with the two layers rotated by $\theta/2$ in opposite directions. The bottom monolayer (sites 1--5, rotated by $-\theta/2$) and the top monolayer (sites 6--10, rotated by $+\theta/2$), with their lattice sites and labels, are shown in blue and red, respectively. The general intralayer hoppings in the moir\'e basis are denoted by $T_{\alpha\beta}^{\mathbf{v}}$ and the interlayer tunneling terms are denoted by $G_{\alpha\beta}^{\mathbf{v}}$ as discussed later. }
\label{36_unitcell}
\end{figure}
The standard model for this system consists of two square lattices, with each lattice site representing a Cu atom. We set the lattice constant of both monolayers to $a=1$. The two layers are twisted by a commensurate angle
$\theta=2\tan^{-1}(m/n)$, where $(m,n)$ are integers that define the twist angle. The corresponding moir\'e unit cell contains $2(m^2+n^2)$ lattice sites and has a lattice constant $b=\sqrt{m^2+n^2}\,a$. Each layer is rotated by $\theta/2$ in opposite directions. Using the $C_4$ symmetry of the system, we map twist angles $\theta>45\degree$ to their equivalent angles $90\degree-\theta$. The moir\'e unit cell for a twist angle of $53.13\degree\equiv36.87\degree$, corresponding to $(m,n)=(1,2)$, is shown in Fig.~\ref{36_unitcell}. This change in notation has two consequences. First, for the orbital-overlap-based tunneling model introduced in Sec.~\ref{sec:tunneling}, the equivalence $\theta\rightarrow 90\degree-\theta$ holds only up to a gauge transformation. Since the $d_{x^2-y^2}$ orbital changes sign under a $90\degree$ rotation, all interlayer tunneling amplitudes change sign, as illustrated by $g_0(53.13\degree)=-g_0(36.87\degree)$ in Fig.~\ref{g0vstheta}. This sign change can be absorbed through the gauge transformation $c_{i\sigma 2}\rightarrow -c_{i\sigma 2}$. Second, the $B_1$ and $B_2$ representations are interchanged under the $45\degree$ rotation relating the two coordinate frames, i.e., $B_1\leftrightarrow B_2$. In our previous work \cite{panda2026}, the order-parameter symmetries at $36.87\degree$ were inadvertently reported using the irreducible representation labels corresponding to the $53.13\degree$ convention, without accounting for this transformation. Consequently, the $B_1$ and $B_2$ labels reported for the $36.87\degree$ twist in Ref.~\cite{panda2026} should be interchanged. Throughout the present work, we refer to these previously reported results using the corrected symmetry labels. This correction does not affect the underlying numerical results.

\par
We solve the two-layer BCS model with nearest-neighbor attraction in each layer,
\begin{equation}
\begin{aligned}
H = & -\sum_{ij,\sigma d} t_{ij} c_{i\sigma d}^\dagger c_{j\sigma d}
- \mu \sum_{i\sigma d} n_{i\sigma d} \\
& - \sum_{ij,d} V_{ij} n_{i d} n_{j d}
- \sum_{ij\sigma} g_{ij} (c_{i\sigma 1}^\dagger c_{j\sigma 2} +\text{H.c.}),
\end{aligned}
\label{eq:Hamiltonian}
\end{equation}
where $i,j$ are the site indices, and $c_{i\sigma d}^\dagger$ and $c_{i\sigma d}$ are the electron creation and annihilation operators, respectively. Here, $d=1,2$ is the monolayer index corresponding to the bottom and top monolayers, respectively. The intralayer hopping term $t_{ij}$ includes nearest-neighbor hopping $t=0.153$~eV and next-nearest-neighbor hopping $t^\prime=-0.45t$, consistent with typical cuprate parameters \cite{Can2021}. All energy units throughout this work are in eV unless otherwise specified. The chemical potential $\mu$ is determined self-consistently to keep the number of electrons (filling) fixed. The band filling is denoted by $\nu$, such that the number of electrons per moir\'e unit cell is equal to $2\nu (m^2+n^2)$. The term $V_{ij}$ represents the nearest-neighbor density-density interaction, which is factorized in the singlet Cooper channel,  leading in the single-layer case to a doping-dependent competition between $d$ and $s$ pairing. The pairing interaction used here is similar to the spin-fluctuation interaction used in Ref.~\cite{Maier2011} and identical to the one employed in Ref.~\cite{Joynt1996}. We use $V=0.2$~eV and temperature $T=10^{-4}$~eV $=1.16$~K throughout this study. The increasing size of the moir\'e unit cells for certain commensurate twist angles near \(45^\circ\) imposes computational constraints on the momentum-grid resolution. We have performed phase diagram calculations for $\theta=0\degree$, $16.26\degree$ ($\equiv73.74\degree$), $22.62\degree$ ($\equiv67.38\degree$), $28.07\degree$, $31.89\degree$, $36.87\degree$ ($\equiv53.13\degree$), $41.11\degree$, and $43.6\degree$. We use momentum grids ranging from $73\times73$ for the $0\degree$ twist to $9\times9$ for the $41.11\degree$ twist, which has 146 sites in the moir\'e unit cell.
\par

We now redefine the Hamiltonian in the moir\'e basis, where $i$ and $j$ are relabeled as $(\alpha,\mathbf{u})$ and $(\beta,\mathbf{v})$, respectively. Here, $\alpha$ and $\beta$ are the site indices within the moir\'e unit cell, and $\mathbf{u}$ and $\mathbf{v}$ are the moir\'e unit-cell coordinates.  Since the system is periodic at commensurate twist angles, we evaluate the hoppings and tunnelings and define the interaction terms between sites in a reference moir\'e unit cell and sites in all other moir\'e unit cells, with the corresponding Fourier factors included when transforming to momentum space. This basis transformation allows us to eliminate the layer indices, as this information is carried by the site indices, i.e., if $\alpha\leq m^2+n^2$, the site belongs to the first monolayer, whereas if $m^2+n^2<\alpha\leq 2(m^2+n^2)$, it belongs to the second monolayer. We have discussed the basis transformation and the algorithm, along with a few examples, in Appendix A of Ref.~\cite{panda2026}. After the basis transformation, within mean-field theory, we solve the SC gap equation self-consistently,
\begin{equation}
    \Delta_{\alpha\beta}(\mathbf k)=\sum_{\mathbf{k}^\prime}V^{\mathbf k \mathbf k^\prime}_{\alpha\beta}\langle c_{\alpha \mathbf k^\prime\up} c_{\beta-\mathbf k^\prime\down} \rangle,
\end{equation}
where
\begin{equation}
V^{\mathbf k \mathbf k^\prime}_{\alpha\beta}=\sum_{\mathbf{r}}V_{\alpha\beta}e^{-i(\mathbf{k}-\mathbf{k}^\prime)\cdot \mathbf{r}},
\end{equation}
$\alpha$ and $\beta$ correspond to the site labels in our moir\'e unit cell, $\mathbf k,\mathbf k'$ are momenta in the moir\'e Brillouin zone, $V_{\alpha\beta}=V$ and $\mathbf{r}=\mathbf{u}-\mathbf{v}$ is the difference between the nearest-neighbor moir\'e unit cell coordinates, since we consider only nearest-neighbor pairing. The symmetry of the twisted bilayer square-lattice system is $D_4$, and hence the possible irreducible representations (irreps), along with their in-plane projections, are $A_1 \rightarrow \text{``}s\text{''} $ wave, $A_2 \rightarrow \text{``}g\text{''} $ wave, $B_1\rightarrow  \text{``}d_{x^2-y^2}\text{''} $ wave, $B_2\rightarrow \text{``}d_{xy}\text{''}$ wave and $E \rightarrow \text{``}p\text{''} $ wave, as listed in Table~I of Ref.~\cite{panda2026}. In the following, we refer to the TRSB $B_1+iB_2$ state as the \mbox{$d+id'$} state; as discussed in Ref.~\cite{panda2026}, this correspondence is only nominal, since the in-plane labels do not capture the full symmetry of the irreps. The size of the order parameter matrix $\Delta_{\alpha\beta}(\mathbf{k})$ increases as we approach $45\degree$. We have explicitly discussed the order parameter matrix structure in real and momentum space in our previous work \cite{panda2026}. We start with a random initial guess for the order parameter and perform the self-consistent iterations until convergence. We then Fourier transform the order parameter onto real space, where we perform a gauge transformation to remove the global phase, enabling us to distinguish true TRSB states from time-reversal-symmetric (TRS) ones. Finally, we project the order parameter onto the basis functions of each irrep of the $D_4$ point group to identify the superconducting order parameter at each point in our phase diagram. We also verify the stability of the converged superconducting state by comparing its free energy with those of competing solutions belonging to different irreps, thereby identifying the energetically favored order-parameter symmetry.
Relative to Ref.~\cite{panda2026}, the model, pairing interaction including its value $V=0.2$~eV, self-consistency, gauge fixing and irrep projection are unchanged. The only change is the interlayer term, Eq.~(\ref{eq:tunnel}) $\rightarrow$ Eq.~(\ref{overlap_integral}), described in Sec.~\ref{sec:tunneling}.

\section{Interlayer tunneling forms}
\label{sec:tunneling}

In this section, we discuss the two models of interlayer tunneling that we compare in our study. The phenomenological tunneling previously used to study these systems assumes that the interlayer tunneling decays exponentially with distance. In contrast, we analyze the tunneling amplitudes and the resulting phase diagram obtained by evaluating the orbital overlap between physically motivated $d_{x^2-y^2}$ wave functions.
\subsection{Phenomenological tunneling}
Initial studies of these systems \cite{Can2021,Senechaltwist2022,Fidrysiak2023, panda2026}  used a phenomenological form of interlayer tunneling, represented by $g_{ij}$, of the form
 \begin{equation}
     g_{ij} = g_0\, e^{-(r_{ij} - c)/\rho},
     \label{eq:tunnel}
 \end{equation}
where $g_0$ is the direct tunneling amplitude, $r_{ij}$ is the three-dimensional distance between sites $i$ and $j$ in different monolayers, $c$ is the interlayer distance, and $\rho$ determines the decay rate of the tunneling.
\par
The TRSB states obtained using this form of tunneling for various $\nu$, $\theta$, and $g_0$ have been studied in detail in Ref.~\cite{panda2026}. In terms of wave function modeling, this form of tunneling is asymptotically realized by the overlap of isotropic $s$-wave wave functions localized at the lattice sites.
The wave function centered at site $\alpha$ has the form
\begin{equation}
    \psi_\alpha(\mathbf{r}) \propto e^{-|\mathbf{r}-\mathbf{r}_\alpha|/a_0},
\end{equation}
where $\mathbf{r}_\alpha$ is the position of lattice site $\alpha$ and $a_0$ is the ``Bohr radius'', which can be varied to control the decay rate of the wave function.
\par
This form provides three tuning parameters that control the tunneling amplitudes. The decay rate $\rho$ determines how rapidly the tunneling amplitude decreases for long-range terms. We choose a value of $\rho=0.3$, similar to previous studies \cite{Can2021}, such that the major contributions arise from the direct, nearest-neighbor, and next-nearest-neighbor tunneling terms. The cutoff used for the orbital-overlap-based tunneling is discussed in Sec.~\ref{Orbital-overlap-based tunneling}. The interlayer distance $c$ and the direct tunneling term $g_0$ both determine the tunneling amplitudes. The effect of changing $c$ can be rescaled by varying $g_0$. In our study of the phenomenological tunneling model, we fix $c=1$ and investigate the phase diagram by varying $g_0$.
\par
The $s$-wave nature of the wave function fails to incorporate any effects of twisting other than the scaling of the moir\'e unit cell. One of the key arguments for explaining the order-of-magnitude reduction in the Josephson critical current at $45\degree$ is the order-parameter mismatch between the two $d_{x^2-y^2}$ order parameters when one monolayer is twisted by $45\degree$ relative to the other. A similar argument can be made for wave function mismatch if the  Wannier wave function on each Cu site is sign changing, which could lead to a  small or vanishing direct tunneling term. The phenomenological tunneling model is incapable of capturing this physics because the wave function does not change sign. Consequently, the direct tunneling term is always the strongest, and all tunneling matrix elements have the same sign. We discuss this in detail when comparing the hopping and tunneling matrices of the two models.

\subsection{Orbital-overlap-based tunneling}
\label{Orbital-overlap-based tunneling}
\begin{figure}[tb]
   \centering
    \includegraphics[width=0.88\linewidth]{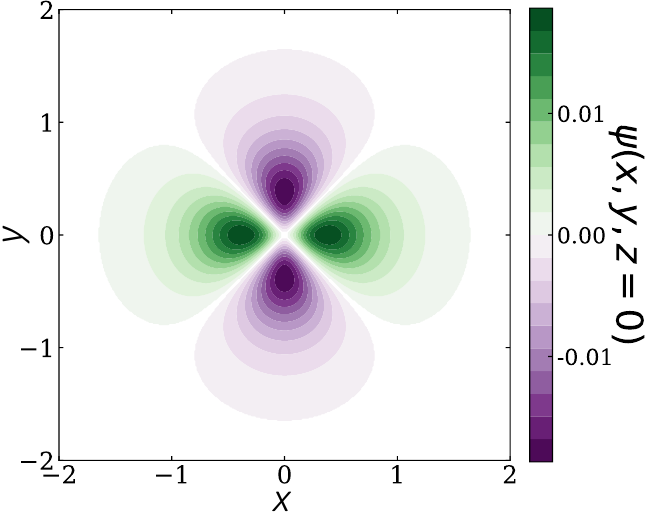}
    \caption{$d_{x^2-y^2}$ wave function for $a_0=1$ and $z=0$. The wave function is later benchmarked to reproduce $t=0.153$~eV and $t^\prime/t=-0.45$. The wave function changes sign under a $90\degree$ rotation and is placed at each site of the square lattice to calculate the orbital overlap with other sites and thereby determine the tunneling amplitudes.}
    \label{dx2y2_wavefunction}
\end{figure}

\begin{figure*}[tb]
   \centering
    \includegraphics[width=0.95\linewidth]{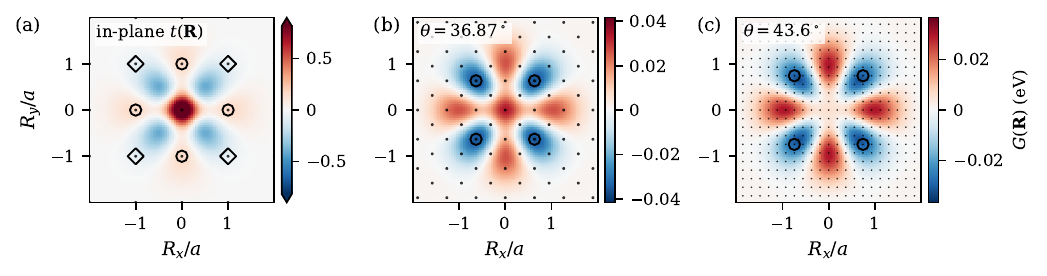}
    \caption{In-plane hopping and interlayer tunneling maps for different twist angles.
(a) In-plane hopping amplitude $t(\mathbf{R})$, the continuum counterpart of $T_{\alpha\beta}^{\mathbf{v}}$ within a monolayer, obtained from the overlap of two benchmarked $d_{x^2-y^2}$ wave functions, one of which is centered at the origin. Open circles (diamonds) mark the NN (NNN) sites, with $t=0.153$~eV ($t^\prime=-0.069$~eV, $t^\prime/t=-0.45$).
(b), (c) Interlayer tunneling amplitude $G(\mathbf{R})$ between a bottom-layer orbital and a top-layer orbital displaced by $(\mathbf{R},c)$, for $c=0.9a$ at $\theta=36.87\degree$ and $43.6\degree$, respectively.
Black dots indicate the in-plane displacements between lattice sites, corresponding to intralayer displacements in (a) and interlayer displacements in (b) and (c). Open circles in (b) and (c) mark the $C_4$-equivalent displacements corresponding to the largest tunneling amplitude in magnitude, $G_{\max}$, within the interlayer cutoffs $r_c=3.19a$ and $3.07a$, respectively. The natural coordinate frame used in panel (b) differs from that of the $(m,n)=(1,2)$ moir\'e unit cell shown in Figs.~\ref{36_unitcell} and \ref{hop-tun_53} by a $45\degree$ rotation.}
    \label{fig:Gmaps}
\end{figure*}

In order to incorporate the sign-changing behavior and symmetries of the low-energy Wannier orbitals, we define a $d_{x^2-y^2}$ wave function at each lattice site. The overlap between these wave functions generates the hopping and tunneling terms. The wave function is of the form
\begin{equation}
    \psi(\mathbf{r})\propto(x^2-y^2)\,e^{-r/a_0},
    \quad
    \psi^{\mathbf{u}}_{\alpha}(\mathbf{r})=\psi\big(R_\alpha(\mathbf{r}-\mathbf{r}^{\mathbf{u}}_\alpha)\big),
    \label{d_wavefunction}
\end{equation}
where $\psi^{\mathbf{u}}_{\alpha}$ is the wave function centered at site $\alpha$ in moir\'e unit cell $\mathbf{u}$, $\mathbf{r}^{\mathbf{u}}_\alpha=(x^{\mathbf{u}}_\alpha,y^{\mathbf{u}}_\alpha,z_\alpha)$ denotes the position of this lattice site, and $R_\alpha$ is the rotation matrix of the monolayer to which site $\alpha$ belongs, $R(+\theta/2)$ for the bottom and $R(-\theta/2)$ for the top monolayer. The vertical coordinate is $z_\alpha=0$ for sites in the bottom monolayer and $z_\alpha=c$ for sites in the top monolayer. 
The wave function for $a_0=1$ is shown in Fig.~\ref{dx2y2_wavefunction}.
For our periodic system, we define a general hopping matrix $T_{\alpha\beta}^{\mathbf{v}}$, which is obtained by evaluating the expectation value of the kinetic energy,
\begin{equation}
    T^{\mathbf{v}}_{\alpha \beta}=-\dfrac{\hbar^2}{2m^\star}\int d^3\mathbf{r}\,\psi_\alpha^{\mathbf{0}}(\mathbf{r})\nabla^2\psi_\beta^{\mathbf{v}}(\mathbf{r}),
    \label{overlap_integral}
\end{equation}
where the rotation matrices $R_\alpha$ and $R_\beta$ enter through Eq.~(\ref{d_wavefunction}). Since the top and bottom monolayers are rotated by $\theta/2$ in opposite directions, this construction incorporates the effect of the twist directly into the wave functions. The general hopping matrix reduces to $+t_{ij}$ when $\alpha$ and $\beta$ belong to the same monolayer, i.e., either $\alpha,\beta\leq m^2+n^2$ or $m^2+n^2<\alpha,\beta\leq 2(m^2+n^2)$. In contrast, when $\alpha$ and $\beta$ belong to different monolayers, i.e., if $\alpha\leq m^2+n^2$, then $m^2+n^2<\beta\leq 2(m^2+n^2)$, and vice versa, the corresponding matrix elements represent the interlayer tunneling terms, denoted by $G_{\alpha\beta}^{\mathbf{v}}$. They correspond to $g_{ij}$ in Eq.~(\ref{eq:Hamiltonian}). We discuss this quantity in detail in later sections to highlight the differences between the phenomenological and orbital-overlap-based tunneling models.

We emphasize that our approach includes only the kinetic-energy contribution, with the phenomenological parameter $m^\star$ chosen to reproduce the relevant hopping amplitudes while retaining the expected angular dependence. This construction should therefore be regarded as an approximate model of the orbital structure rather than a true Wannier representation. A complete Wannier description would also incorporate  nearby atomic wavefunction and the underlying crystal potential, and Wannier functions centered at different lattice sites would form an orthonormal basis.

In order to obtain the normal-state Hamiltonian parameters relevant for BSCCO, we first benchmark the in-plane hopping amplitudes on a square lattice.
Tuning the ratio $a_0/a$ fixes the ratio $t'/t$; examining the nearest-neighbor (NN) $(x,0)$ and next-nearest-neighbor (NNN) $(x,x)$ hoppings as a function of $a_0/a$, we find $a_0=0.1848a$. 
 The overall magnitude $t=0.153$~eV is then fixed by $m^\star$. The in-plane hopping map is shown in Fig.~\ref{fig:Gmaps}(a). The on-site term only shifts the chemical potential and is therefore absorbed into $\mu$. The NN and NNN hopping amplitudes have opposite signs, and the overlap decays on the scale of the unit-cell lattice vector.

\par
 In contrast to the phenomenological tunneling model, the effects of twisting are incorporated through the rotation of the wave function itself, in addition to the scaling of the moir\'e unit cell. The $d_{x^2-y^2}$ wave functions in the top and bottom monolayers are rotated in opposite directions, leading to different overlap amplitudes at different twist angles. Consequently, the interlayer tunneling matrix elements can have both positive and negative signs depending on which lobes overlap most strongly in the integral. The cutoff distance for evaluating the tunneling amplitudes is defined as the distance after which the tunneling amplitudes are less than $1\%$ of the maximum tunneling amplitude $G_{\max}$.
The interlayer tunneling amplitudes $G(\bf R)$, the continuum counterpart of $G_{\alpha\beta}^{\mathbf{v}}$, for $\theta=36.87\degree$ (in the natural coordinate frame) and $43.6\degree$ are shown in Figs.~\ref{fig:Gmaps}(b) and \ref{fig:Gmaps}(c), respectively. The amplitudes correspond to tunneling from the site at $(0,0)$ in the bottom monolayer to sites at positions $(\mathbf{R},c)$ in the top monolayer. The sites in the top monolayer are represented by black dots, while those corresponding to $G_{\max}$ are marked by open circles. The tunneling maps exhibit sign-changing amplitudes, with long-range tunneling terms dominating over the direct tunneling term at both twist angles.

\begin{figure}[tb]
   \centering
    \includegraphics[width=0.95\linewidth]{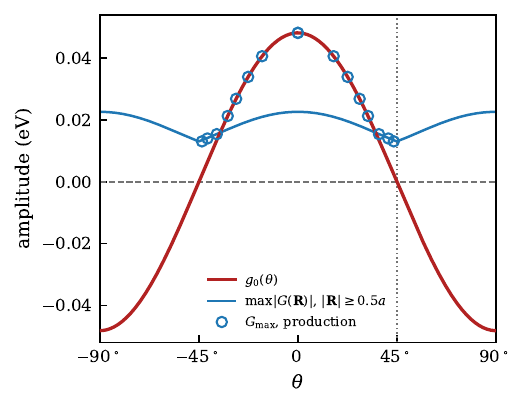}
    \caption{Direct tunneling amplitude $g_0(\theta)$ (red) and the largest off-vertical tunneling amplitude $\max|G(\mathbf{R})|$ over in-plane displacements $|\mathbf{R}|\geq0.5a$ (blue) for the orbital-overlap-based model with $c=1.1a$. The orbital overlap changes as the wave functions in the two layers are rotated by $\theta/2$ in opposite directions, resulting in $g_0\propto\cos2\theta$. At $45\degree$ (dotted line), the positive and negative contributions to the direct orbital overlap cancel, resulting in $g_0=0$. The blue curve is evaluated in the continuum, independently of the lattice geometry, with its maximum occurring at $|\mathbf{R}|\approx0.84$--$0.98a$. The cusps at $\pm45\degree$ reflect the exchange of two equivalent lobes. The open circles represent the largest tunneling amplitudes $G_{\max}$ obtained from the tunneling matrices used in our calculations (with cutoff $r_c$, as discussed in Sec.~\ref{sec:matrices}) at the commensurate twist angles $\pm\theta$, where $\theta=0\degree$, $16.26\degree$, $22.62\degree$, $28.07\degree$, $31.89\degree$, $36.87\degree$, $41.11\degree$, and $43.6\degree$.}
    \label{g0vstheta}
\end{figure}

To further investigate this behavior, we compare the direct tunneling amplitude with those of the longer-range tunneling terms. Incorporating the twist directly into the wave functions leads to a strong twist-angle dependence of the direct tunneling amplitude, as shown in Fig.~\ref{g0vstheta}. The values of $g_0$ at $0\degree$ and $90\degree$ have the same magnitude but opposite signs due to the overlap between lobes of opposite signs when the two layers are rotated by $45\degree$ in opposite directions. This angular dependence provides a consistency check for our construction: since the Laplacian is rotationally invariant, it preserves the rotational transformation properties of the $d_{x^2-y^2}$ wave function, and the resulting overlap therefore retains the expected $\cos{2\theta}$ angular dependence. At $\theta=45\degree$, $g_0=0$ because the positive and negative contributions to the orbital overlap cancel. The off-vertical tunneling terms are nonzero and sign-changing at every twist angle, and their largest magnitude depends only weakly on $\theta$ (blue line in Fig.~\ref{g0vstheta}). $G_{\max}=g_0$ up to $\theta=31.89\degree$, whereas at $36.87\degree$, $41.11\degree$, and $43.6\degree$ the off-vertical terms dominate, with $G_{\max}/|g_0|=1.14$, $2.15$, and $5.57$ for $c=1.1a$; $G_{\max}$ remains at $13$--$15$~meV while $g_0\rightarrow0$ at $45\degree$.

The direct tunneling amplitude is not an input parameter in this model but can be varied by changing the $c$-axis distance. It is important to note that the $c$-axis distance in this calculation can be varied over a relatively large range because we are not using the exact cuprate Wannier wave function. We later vary the $c$-axis distance to generate a phase diagram while generally ensuring that $g_0 \ll t$. We vary the interlayer distance over a range $c\in[0.65,1.5]$ (in units of $a$), with the exact range depending on the convergence of the self-consistent calculations, particularly for small $G_{\max}$. Since the orbitals are model functions, $c$ should be regarded as an effective parameter that controls the interlayer coupling rather than as the actual CuO$_2$--CuO$_2$ distance across the twisted interface, which is considerably larger. In general, this orbital-overlap-based tunneling model allows us to capture several effects, such as sign changes in the tunneling matrix elements and larger off-vertical tunneling terms, which lead to considerable differences in the resulting Fermi surface. 

\section{Hopping and tunneling matrices}
\label{sec:matrices}
As the twist angle approaches $45\degree$, the number of sites in the moir\'e unit cell increases rapidly, from 2 sites at $0\degree$ to, for example, 58 sites at $43.6\degree$. This results in increasingly large general hopping matrices. To investigate the differences between the phenomenological and orbital-overlap-based tunneling models, we compare the general hopping matrices $T_{\alpha\beta}^{\mathbf{v}}$ at various twist angles.
\par

\begin{figure}[tb]
   \centering
\includegraphics[width=0.94\linewidth]{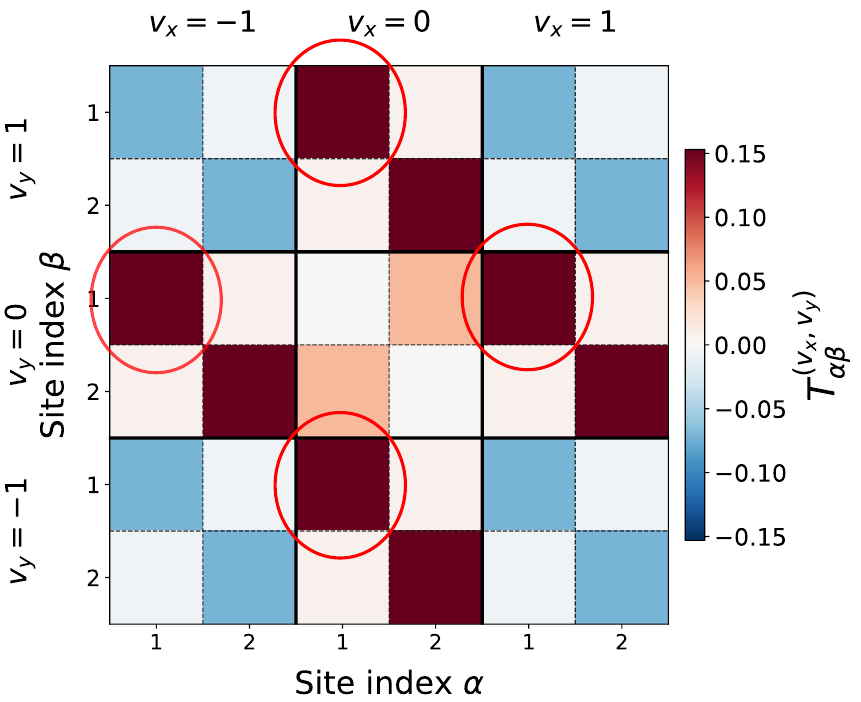}
    \caption{General hopping matrix $T_{\alpha\beta}^{\mathbf{v}}$ plotted for $\theta=0\degree$ and $G_{\max}=g_0\approx0.048$ ($c=1.1$) using the $d_{x^2-y^2}$ wave-function overlap. Site $\alpha$ represents a site in the $(0,0)$ moir\'e unit cell, whereas the site index $\beta$ represents a site in the $(v_x,v_y)$ moir\'e unit cell. For a $0\degree$ twist, since there is one site per monolayer in the unit cell, the nearest neighbors of site 1 are located in the nearest-neighbor moir\'e unit cells. The matrix elements corresponding to these nearest neighbors are marked by red circles.}
    \label{0_hop_tun_orbital}
\end{figure}
\par
The intralayer hopping is restricted to NN and NNN hopping. For the interlayer tunneling, we include all pairs of sites in different monolayers with in-plane distance $|\mathbf{R}|\le r_c$, with $r_c=3.07$--$3.38a$ depending on the twist angle. In the matrix plots below, we show the matrix elements corresponding to the same moir\'e unit cell $(0,0)$, the nearest-neighbor (NN) moir\'e unit cells, e.g., $(0,1)$ and $(-1,0)$, and the next-nearest-neighbor (NNN) moir\'e unit cells, e.g., $(1,1)$ and $(-1,1)$. Although our Hamiltonian includes longer-range tunneling terms, we restrict the matrix plots to these unit cells for better visualization. We can understand the resulting matrix structure by considering the general hopping matrix for a $0\degree$ twist shown in Fig.~\ref{0_hop_tun_orbital}, generated using the orbital-overlap-based tunneling model for $G_{\max}=g_0\approx0.048$ ($c=1.1$). There are two sites in the moir\'e unit cell, one from each monolayer, located directly on top of each other. $T_{\alpha \beta}^{\mathbf{v}}$ is a $6\times6$ matrix comprising nine $2\times2$ matrices, with each of the nine blocks corresponding to a particular moir\'e unit cell. The $2\times2$ matrix associated with each moir\'e unit cell $\mathbf{v}$ contains the hopping and tunneling amplitudes between site $\alpha$ in the reference unit cell and site $\beta$ in unit cell $\mathbf{v}$.
\par
\begin{figure}[tb]
    \centering
    \includegraphics[width=0.98\linewidth]{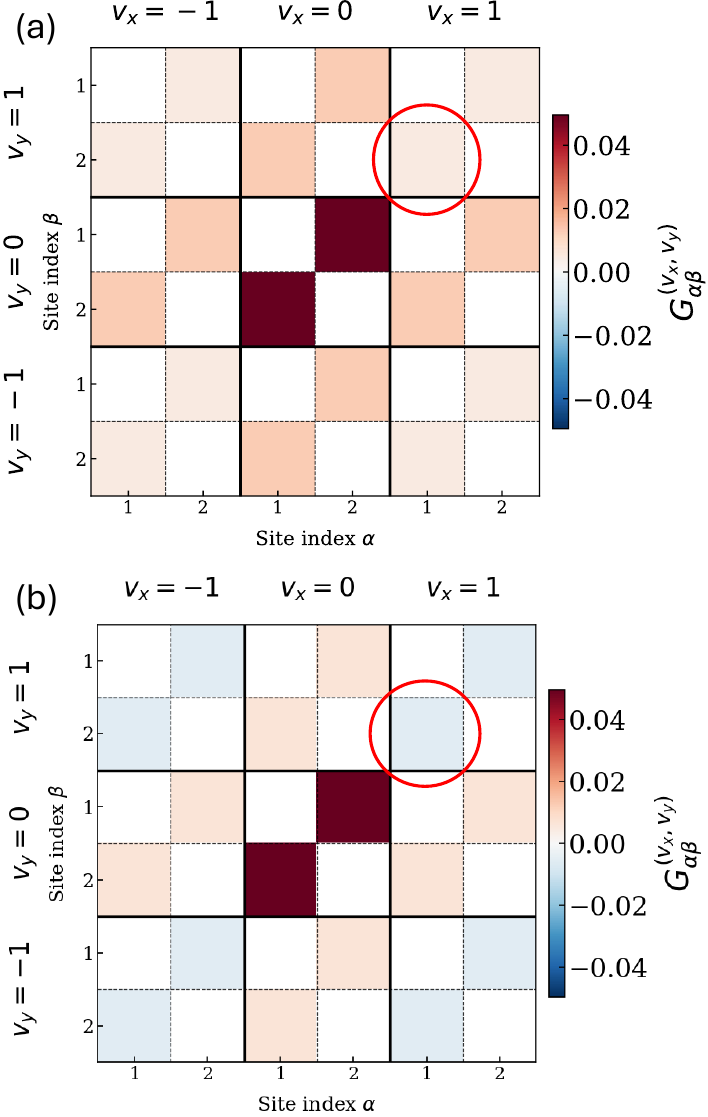}
    \caption{Comparison of the tunneling amplitudes
    from the phenomenological tunneling model and the orbital-overlap-based tunneling model at $0\degree$ for the same set of parameters as in Fig.~\ref{0_hop_tun_orbital}. (a) $G_{\alpha\beta}^{\mathbf{v}}$ for the phenomenological tunneling model, showing the decay of the tunneling amplitude for long-range tunneling terms. Since this form can be modeled using an isotropic $s$-wave wave function, there are no sign changes in the tunneling amplitudes. (b) $G_{\alpha\beta}^{\mathbf{v}}$ obtained from the orbital-overlap-based tunneling model. At $0\degree$, the direct tunneling term has the largest amplitude, while the sign-changing wave function generates sign changes in the tunneling amplitudes, as indicated by the blue regions of the matrix. We mark the $G_{12}^{(1,1)}$ matrix element with red circles in both panels to highlight the sign difference between the two models. The corresponding parameter point is marked by a black circle in the $0\degree$ panel of Fig.~\ref{Phase diagram}.}
    \label{0_twist_tun}
\end{figure}
\begin{figure}[tb]
   \centering
    \includegraphics[width=0.95\linewidth]{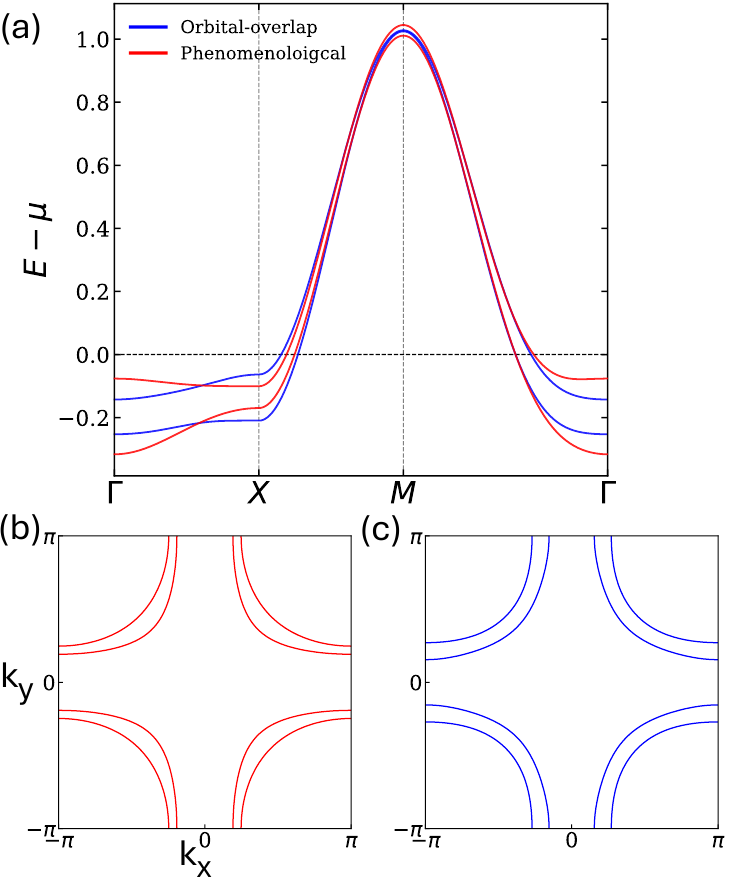}
    \caption{Band structure and Fermi surfaces obtained from the two tunneling models for $\nu=0.49$, $G_{\max}=g_0\approx0.048$ ($c=1.1$), and $\theta=0\degree$.  (a) Band structure for the two models plotted along the high-symmetry path $\Gamma \rightarrow X \rightarrow M \rightarrow \Gamma$. The phenomenological tunneling model is shown in red and the orbital-overlap-based tunneling model in blue. (b) Phenomenological tunneling model with a $B_1+iA_1$ ground state. (c) Orbital-overlap-based tunneling model with a $B_1$ ground state. Although the direct tunneling term is the dominant tunneling amplitude in both cases, the sign-changing tunneling amplitudes in the orbital-overlap-based model lead to visible differences in the band structure and Fermi surface and stabilize a different SC ground state.}
    \label{FS_0}
\end{figure}
For the $0\degree$ case, since there is only one site per monolayer in the moir\'e unit cell, there are no intra-unit-cell nearest-neighbor hopping terms, and hence \mbox{$T_{11}^{(0,0)}=T_{22}^{(0,0)}=0$}. However, there is a direct tunneling term between sites 1 and 2, and hence \mbox{$T_{12}^{(0,0)}=T_{21}^{(0,0)}=g_0$}. Electrons on site 1 in the $(0,0)$ unit cell  undergo nearest-neighbor hopping to  site 1 in the $(0,1)$, $(0,-1)$, $(1,0)$, and $(-1,0)$ unit cells. These correspond to the matrix elements $T_{11}^{(0,1)}=T_{11}^{(0,-1)}=T_{11}^{(1,0)}=T_{11}^{(-1,0)}=t$, which are marked by red circles in Fig.~\ref{0_hop_tun_orbital}.
\begin{figure*}[tb]
    \centering
    \includegraphics[width=0.95\linewidth]{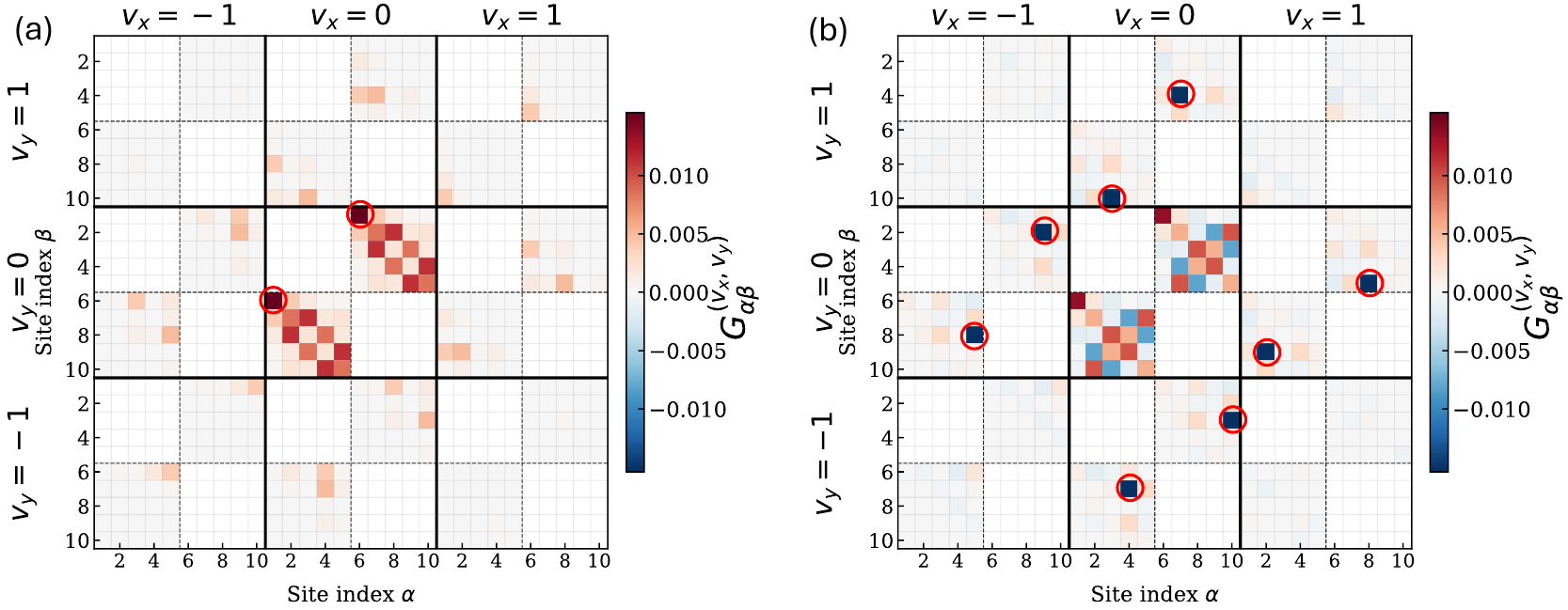}
    \caption{Comparison of the tunneling matrices obtained from the two models for $\theta=36.87\degree$. The off-diagonal blocks containing the interlayer tunneling amplitudes are highlighted using gray shading. The moir\'e unit cell contains 10 sites, as shown in Fig.~\ref{36_unitcell}, with 5 sites in each monolayer. (a) $G_{\alpha\beta}^{\mathbf{v}}$ for the phenomenological tunneling model with $g_0\approx0.0154$. The tunneling matrix exhibits no sign changes, with the maximum amplitude corresponding to $G_{16}^{(0,0)}=G_{61}^{(0,0)}$ (marked in red), which represents the direct tunneling term. All other longer-range tunneling terms have smaller amplitudes. (b) $G_{\alpha\beta}^{\mathbf{v}}$ for the orbital-overlap-based tunneling model with $G_{\max}\approx0.0154$. The tunneling matrix exhibits sign-changing amplitudes and off-vertical tunneling terms with larger amplitudes than the direct tunneling term. The largest tunneling amplitudes are marked by red circles and correspond to $G_{4,10}^{(0,-1)}=G_{9,2}^{(1,0)}=G_{3,7}^{(0,1)}=G_{5,8}^{(1,0)}$ and their Hermitian conjugates. The corresponding parameter point is marked by a black circle in the $36.87\degree$ panel of Fig.~\ref{Phase diagram}.}
    \label{hop-tun_53}
\end{figure*}
In general, for any commensurate twist angle $(m,n)$, $T_{\alpha\beta}^{\mathbf{v}}$ is a $6(m^2+n^2)\times6(m^2+n^2)$  matrix consisting of nine \mbox{$2(m^2+n^2)\times2(m^2+n^2)$} blocks. Within each block, the first $(m^2+n^2)$ indices belong to the first monolayer, while the remaining indices belong to the second monolayer. In each \mbox{$2(m^2+n^2)\times2(m^2+n^2)$} block, the two diagonal \mbox{$(m^2+n^2)\times(m^2+n^2)$} blocks contain the intralayer hopping amplitudes, while the two off-diagonal \mbox{$(m^2+n^2)\times(m^2+n^2)$} blocks contain the interlayer tunneling amplitudes.
\par

In our previous study of the phase diagram of the twisted bilayer cuprate system, we also investigated regimes with very strong interlayer tunneling. However, since cuprates are van der Waals materials, the interlayer tunneling amplitudes (between the bilayers in BSCCO) are expected to be much smaller than the intralayer hopping amplitudes. We have also discussed how the twist angle affects the interlayer tunneling amplitudes, with the direct tunneling term vanishing as the twist angle approaches $45\degree$. Since we work primarily in the $G_{\max}\ll t$ limit, we evaluate the interlayer tunneling matrix $G_{\alpha\beta}^{\mathbf{v}}$, for various twist angles in order to isolate and visualize the changes in the tunneling matrix elements. Since $G_{\alpha\beta}^{\mathbf{v}}$ contains only tunneling terms, all \mbox{$(m^2+n^2)\times(m^2+n^2)$} diagonal blocks are zero in each of the nine moir\'e unit-cell blocks.

We show the full tunneling matrix for $\theta=0\degree$ and $G_{\max}=g_0\approx0.048$ ($c=1.1$) in Fig.~\ref{0_twist_tun}. Figure~\ref{0_twist_tun}(a) displays the tunneling matrix for the phenomenological model. The largest amplitude corresponds to the direct tunneling term $G_{12}^{(0,0)}=G_{21}^{(0,0)}$. All other tunneling terms have the same sign because the isotropic $s$-wave wave function does not allow for sign-changing tunneling amplitudes. In Fig.~\ref{0_twist_tun}(b), we show the tunneling matrix obtained from the orbital-overlap-based tunneling model. The dominant term remains $G_{12}^{(0,0)}=G_{21}^{(0,0)}$, but in this case we clearly observe sign changes in the tunneling matrix elements. As an example, we compare $G_{12}^{(1,1)}$ for the two models, as indicated by the red circles.
\par

We demonstrated in our previous work \cite{panda2026} how changes in the Fermi surface redistribute spectral weight across different regions of the Brillouin zone and thereby favor different order-parameter irreps. Similar effects can be observed for $\theta=0\degree$. Although the direct tunneling term remains the largest amplitude in the interlayer tunneling matrix $G_{\alpha\beta}^{\mathbf{v}}$ for both models, the sign-changing tunneling amplitudes in the orbital-overlap-based model lead to differences in the band structures and Fermi surfaces. These differences are illustrated in Fig.~\ref{FS_0} for the same filling, $\nu=0.49$, and maximum tunneling amplitude $G_{\max}=g_0\approx0.048$ ($c=1.1$). Figure~\ref{FS_0}(a) compares the band structures obtained from the two models along the high-symmetry path $\Gamma \rightarrow X \rightarrow M \rightarrow \Gamma$, with the phenomenological tunneling model shown in red and the orbital-overlap-based tunneling model shown in blue. The two models exhibit distinct band splittings along the $X\rightarrow M$ path. In the orbital-overlap-based model, the band splitting decreases as the $M$ point is approached and becomes nearly negligible in its vicinity. Figure~\ref{FS_0}(b) shows the Fermi surface obtained from the phenomenological tunneling model, which yields a $B_1+iA_1$ ground state, consistent with our previous work \cite{panda2026}. In contrast, Fig.~\ref{FS_0}(c) shows the Fermi surface obtained from the orbital-overlap-based tunneling model, which yields a $B_1$ ground state, as also shown in the phase diagram discussed below.
\par

\begin{figure}[tb]
   \centering
   \includegraphics[width=0.88\linewidth]{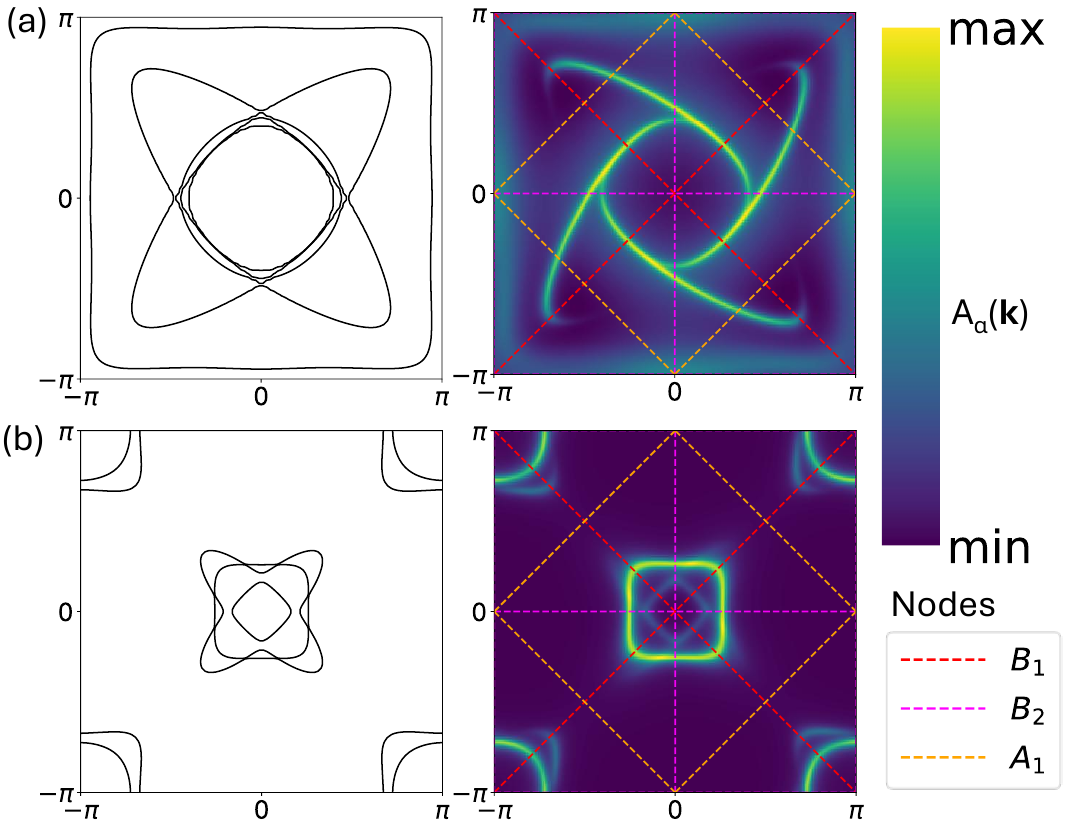}
   \caption{Fermi surfaces and spectral functions obtained from the two tunneling models for $\nu=0.4$, $G_{\max}=0.076$ ($c=0.7$), and $\theta=36.87\degree$ ($\equiv53.13\degree$). The calculations use the $53.13\degree$ coordinate convention, with the nodal lines of the different irreps shown in the legend to identify regions of large spectral weight near their corresponding antinodes. The $B_1\leftrightarrow B_2$ transformation is applied when identifying the irreps for the equivalent $36.87\degree$ twist. The spectral function is shown for site~1, which lies on the $C_4$ rotation axis passing through site~6 in the top monolayer (Fig.~\ref{36_unitcell}). (a) The phenomenological tunneling model exhibits substantial spectral weight near the $\Gamma$ and $M$ points, corresponding to the antinodes of the $A_1$ and $B_1$ irreps, respectively, and yields a $B_1+iA_1$ ground state. (b) The orbital-overlap-based model exhibits spectral weight primarily near the $B_1$ antinodes, with additional weight near $\Gamma$ suggesting a competing $A_1$-wave instability. Free-energy comparisons among the $B_1$, $A_1$, and $B_1+iA_1$ solutions confirm that the $B_1$ state is energetically favored. The corresponding parameter point is marked by a green circle in the $36.87\degree$ panel of Fig.~\ref{Phase diagram}.}
   \label{FS_53}
\end{figure}
Further insight into the effects of twisting can be obtained by examining the tunneling matrix $G_{\alpha\beta}^{\mathbf{v}}$ at \mbox{$\theta=36.87\degree$} and $G_{\max}\approx0.0154$  ($c=1.1$). The direct tunneling amplitude for the same $c$-axis distance is different from the $0\degree$ case due to the angle dependence of the direct tunneling term shown in Fig.~\ref{g0vstheta}. We have nine blocks representing the nine moir\'e unit cells, with each block being a $10\times10$ matrix since there are 10 sites in the moir\'e unit cell, as shown in Fig.~\ref{36_unitcell}. The two diagonal $5\times5$ blocks are zero for every moir\'e unit-cell coordinate, while the off-diagonal $5\times5$ blocks contain the tunneling amplitudes.
\par

Figure~\ref{hop-tun_53}(a) shows the tunneling matrix for the phenomenological tunneling model for $g_0=0.0154$, which exhibits features similar to those observed at the $0\degree$ twist angle. The strongest term is the direct tunneling term $G_{16}^{(0,0)}=G_{61}^{(0,0)}$, marked by red circles, and all the tunneling amplitudes have the same sign. The tunneling matrix generated by the orbital-overlap-based tunneling model for $G_{\max}=0.0154$ ($g_0=0.0135$)  is shown in Fig.~\ref{hop-tun_53}(b). In contrast to the phenomenological model, it exhibits sign-changing tunneling amplitudes as well as large off-vertical tunneling terms. We find that the largest interlayer tunneling terms are $G_{4,10}^{(0,-1)}=G_{9,2}^{(1,0)}=G_{3,7}^{(0,1)}=G_{5,8}^{(1,0)}$ and their Hermitian conjugates, as indicated by the red circles.
\par

 The normal-state sublattice spectral function is defined by the diagonal components of the full spectral function matrix
\begin{align}
    A_{\alpha}(\mathbf{k},\omega)=-\dfrac{1}{\pi}\mathop{\text{Im}}[(\omega+i\eta)\mathbbm{1}-H_0(\mathbf k)]^{-1}_{\alpha\alpha} ,
\end{align}
where $H_0$ is the noninteracting part of the Hamiltonian as derived in the appendix of our previous work \cite{panda2026}, $\mathbbm{1}$ is the identity matrix, and $\alpha$ is the moir\'e site index. For the Fermi surface, $\omega=0$, and we examine $A_{\alpha}(\mathbf{k})$ to investigate the presence of high spectral weight in the Fermi surface antinodal regions  of a given order parameter symmetry since such spectral weight favors the state by lowering the condensation energy.
\par
The differences in $G_{\alpha\beta}^{\mathbf{v}}$ lead to considerable changes in the Fermi surfaces and spectral functions, as shown in Fig.~\ref{FS_53} for $\nu=0.4$, $G_{\max}=0.076$ ($c=0.7$) and $\theta=36.87\degree\equiv 53.13\degree$. The Fermi surfaces and spectral functions are calculated using the $53.13\degree$ coordinate convention, and the $B_1\leftrightarrow B_2$ transformation is applied when identifying the order-parameter symmetries for the equivalent $36.87\degree$ twist. Figure~\ref{FS_53}(a), obtained using the phenomenological tunneling model, yields a $B_1+iA_1$ ground state. In this case, there is substantial spectral weight near the $\Gamma$ and $M$ points, which correspond to the antinodes of the $A_1$ and $B_1$ irreps, respectively. Figure~\ref{FS_53}(b), obtained using the orbital-overlap-based tunneling model, displays a substantially different Fermi surface, with the spectral weight concentrated near the antinodes of the $B_1$ irrep. However, we also observe spectral weight near the $\Gamma$ point, suggesting a competing $A_1$-wave instability. Despite this competition, the superconducting ground state obtained at this parameter point has $B_1$ symmetry. We have verified this result by comparing the free energies of the competing $B_1$, $A_1$, and $B_1+iA_1$ solutions, confirming that the $B_1$ state is energetically favored.
\par

The observed differences in the normal-state properties motivate a thorough investigation of the phase diagram of these twisted bilayer cuprate systems using the orbital-overlap-based tunneling model. We perform a scan of the converged superconducting order parameter by varying the filling $\nu$ and the largest interlayer tunneling amplitude $G_{\max}$, which is controlled by the interlayer distance $c$.
\section{Phase diagram}
\begin{figure*}[tb]
 \includegraphics[width=\linewidth]{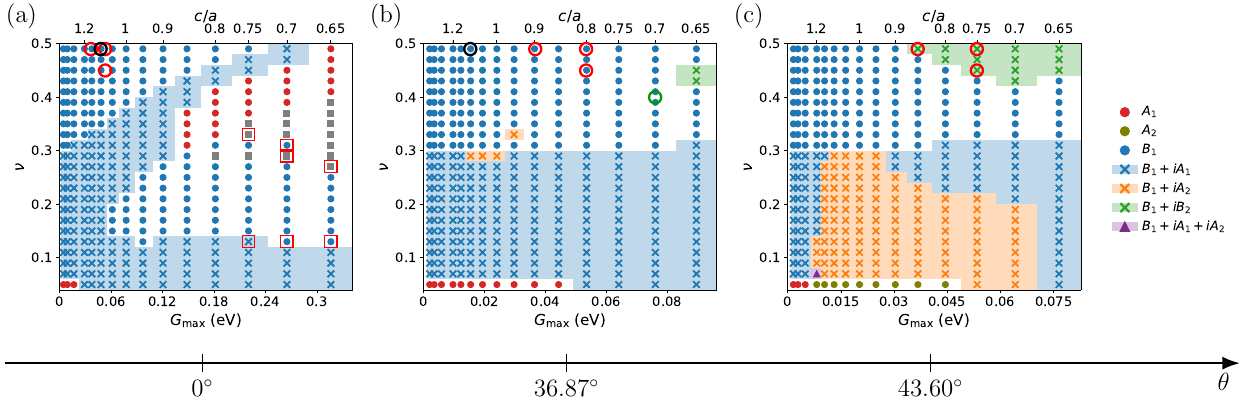}
    \caption{Phase diagram of the twisted bilayer cuprate system obtained from the orbital-overlap-based tunneling model for twist angles (a) $0\degree$, (b) $36.87\degree$, and (c) $43.6\degree$. We determine the SC ground state as a function of the filling $\nu\in[0.05,0.49]$ and the interlayer distance $c\in[0.65a,1.5a]$, using $a_0=0.1848a$ and the interlayer cutoff $r_c$ of Sec.~\ref{sec:matrices}. The bottom axis gives the largest interlayer tunneling amplitude $G_{\max}$ of the Hamiltonian (Fig.~\ref{g0vstheta}), while the top axis shows the corresponding $c/a$. Dots denote TRS states: $A_1$ (red), $A_2$ (olive), and $B_1$ (blue). Crosses denote TRSB states, with the shaded regions indicating the corresponding phases: $B_1+iA_1$ (blue), $B_1+iA_2$ (orange), and $B_1+iB_2$ (green). Purple triangles mark $B_1+iA_1+iA_2$ states. Colored circles indicate the parameter locations of calculations presented in other figures: black circles correspond to the tunneling matrices in Figs.~\ref{0_twist_tun} ($0\degree$) and \ref{hop-tun_53} ($36.87\degree$), the green circle to the spectral functions in Fig.~\ref{FS_53}, and the red circles to the nine parameter points used for the critical current $j_c$ in Fig.~\ref{jc_fixed_gmax_sc}. These circles mark the parameter values of the corresponding calculations and do not necessarily coincide with the sampled points used to construct the phase diagram.}
    \label{Phase diagram}
\end{figure*}

The phase diagram of twisted bilayer cuprates depends strongly on the normal-state properties, and hence different forms of interlayer tunneling are expected to modify the resulting phase diagram. In our previous work \cite{panda2026}, we calculated the phase diagram using the phenomenological tunneling model. The corresponding phase diagrams obtained using the orbital-overlap-based tunneling model are shown in Fig.~\ref{Phase diagram}. We have performed calculations for $\theta=0\degree$, $16.26\degree$ ($\equiv73.74\degree$), $22.62\degree$ ($\equiv67.38\degree$), $28.07\degree$, $31.89\degree$, $36.87\degree$ ($\equiv53.13\degree$), $41.11\degree$, and $43.6\degree$. In the main text, we focus on the phase diagrams for $0\degree$, $36.87\degree$, and $43.6\degree$ to facilitate comparison with our previous work and highlight the key differences between the two tunneling models. The phase diagrams for the remaining commensurate twist angles are presented in Appendix~\ref{Phase_diagram_rest}.

The phase diagram for the $0\degree$ twist angle shows the superconducting ground states for $G_{\max}\in[0.004,0.316]$ and $\nu\in[0.05,0.49]$. In our previous work, we evaluated the phase diagram using phenomenological tunneling for $g_0\in[0,0.3]$ to capture the various phase boundaries. However, since cuprates are van der Waals materials, the interlayer tunneling amplitudes are expected to be weaker than the in-plane hopping amplitudes. For the $0\degree$ phase diagram, we extend our calculations to a range of tunneling strengths comparable to that considered in the phenomenological model to investigate the differences between the two tunneling models, whereas for twist angles closer to $45\degree$, we restrict our analysis to the $G_{\max}\ll t$ regime.

Similar to the phenomenological phase diagram, we find a TRSB $B_1+iA_1$ ($d_{x^2-y^2}+is$-wave in plane) state between the $A_1$-wave state ($s$-wave in plane) at low filling and the $B_1$-wave  ($d_{x^2-y^2}$-wave in plane) states closer to half filling for small $G_{\max}$. As $G_{\max}$ increases, the $B_1+iA_1$ phase bifurcates into two regions, one at small $\nu$ and the other closer to half filling. These two regions are separated by TRS $B_1$-wave and $A_1$-wave states. This is in contrast to the phenomenological phase diagram, where the $B_1+iA_1$ region forms a continuous phase extending from smaller $\nu$ toward half filling and eventually vanishes with increasing $g_0$. In the $G_{\max}=g_0\rightarrow0$ limit, corresponding to decoupled monolayers, we recover the results previously obtained for the monolayer system \cite{Joynt1996,Breio2022}. In particular, the superconducting ground state undergoes a transition from an $A_1$-wave state to a $B_1+iA_1$ state and eventually to a pure $B_1$-wave state as the filling increases. At higher tunneling strengths, we find a few points in the phase diagram, marked by gray squares, for which no superconducting solution was obtained. 
\par
The $36.87\degree$ twist angle was the initial commensurate angle at which a TRSB $B_1+iB_2$ ($d_{x^2-y^2}+id_{xy}$-wave in plane) state was predicted in previous studies with phenomenological tunneling; 
we also found this state at weak tunneling amplitudes within our phenomenological model. 
In the phenomenological phase diagram, we found a continuous region of the $B_1+iB_2$ phase that persisted to larger $g_0$ as the filling approached half filling. The $B_1+iB_2$ phase was in proximity to a $B_1+iA_1$ ($d_{x^2-y^2}+is$-wave in plane)  phase, with the two separated by a narrow region of the TRS $B_1$-wave state.
\par
Probing the small-$G_{\max}$ regime of the phase diagram, $G_{\max}\in[0.001,0.09]$, using the orbital-overlap-based tunneling model shows the complete disappearance of the $B_1+iB_2$ state within the range studied. Instead, we find a TRSB $B_1+iA_1$ state at low fillings and a pure $B_1$-wave state closer to half filling. At higher fillings, the TRSB $B_1+iB_2$ and $B_1+iA_1$ states are largely absent from the phase diagram, except within a very narrow range of fillings at $c=0.65$. We find a continuous region of $B_1+iA_1$ states at lower fillings, with pure $A_1$-wave states appearing at very low fillings and weak tunneling amplitudes. This behavior is consistent with the $G_{\max}\rightarrow0$ limit, as seen for $0\degree$ twist, where the monolayers become decoupled and the relative twist angle has no effect on the superconducting ground state. The differences in the Fermi surfaces and spectral weights between the two models, as shown in Fig.~\ref{FS_53}, indicate a redistribution of spectral weight, thereby stabilizing different order parameters. Consequently, the two tunneling models yield different SC ground states for the same $\nu$ and $G_{\max}$.
\par
For the phenomenological model, the closest angle to $45\degree$ for which we calculated the phase diagram was $43.6\degree$. We previously observed a $B_1+iB_2$ phase at small tunneling strengths, which transitioned into a $B_1+iA_1$ phase with increasing $g_0$ and eventually into an $A_1$-wave phase.  In the present calculation, we restrict our analysis to the $G_{\max}\ll t$ regime and observe a region of $B_1+iB_2$ states close to half filling for $G_{\max}\in[0.038,0.076]$. We also find a $B_1+iA_1$ and $B_1+iA_2$ phase at lower fillings and at very small fillings observe $A_1$ and $A_2$ states. It is important to note that, since there are 58 sites in the moir\'e unit cell, these calculations are computationally expensive and are therefore performed using a relatively sparse momentum grid. We have evaluated the free energy associated with each irrep and verified that the states shown in the phase diagram correspond to the energetically favored solutions rather than artifacts of the initial guesses used in the self-consistent calculations.

In contrast to the phase diagram reported for the phenomenological tunneling model in Ref.~\cite{panda2026}, we find no $B_1+iB_2$ state at $43.6\degree$ for the weakest tunneling amplitudes considered here. Instead, the $B_1+iB_2$ region appears only above a finite $G_{\max}$ and grows with increasing $G_{\max}$, as shown in Fig.~\ref{Phase diagram}. We note, however, that in the decoupled-monolayer limit ($G_{\max}=g_0=0$), the relative phase between the superconducting order parameters of the two layers is undetermined. In our previous work \cite{panda2026}, computational limitations prevented us from investigating tunneling amplitudes as small as those accessible in the present orbital-overlap-based model. Consequently, the $B_1+iB_2$ region was extended from the smallest tunneling amplitude for which a converged solution was obtained down to $g_0=0$. The present results suggest that in the more realistic orbital based model, a  critical value of the interlayer tunneling strength may be necessary to stabilize a well-defined $B_1+iB_2$ state.

\par
The overarching observation is a substantial reduction in the region of the phase diagram where we had previously predicted a $B_1+iB_2$ ground state. Most importantly, we find a very small region of $B_1+iB_2$ state at a $36.87\degree$ twist, where such a state has previously been predicted by a variety of continuum and lattice models for a large range of tunneling amplitudes. This emphasizes the need to develop a comprehensive microscopic model that incorporates the relevant cuprate physics, the symmetry of the system, and, ultimately, the effects of neighboring atoms in the BSCCO monolayers.
\section{Josephson critical current}
Our previous work demonstrated that the strongly competing phases in twisted bilayer cuprate systems can lead to different twist-angle dependencies of the Josephson critical current $J_c$ as the parameters of the Hamiltonian are varied. For smaller tunneling amplitudes $g_0$, we found that $J_c$ decreased with increasing twist angle and was up to two orders of magnitude smaller at $43.6\degree$ than at $0\degree$. In contrast, for larger $g_0$, we did not obtain a $B_1+iB_2$ state, and $J_c$ remained large throughout the range of $\theta$ considered. The expression for the Josephson current is
 \begin{equation}
    J(\phi)=-\dfrac{ie}{\hbar}\sum_{(\alpha,\mathbf{0}),(\beta,\mathbf{v})\sigma}G_{\alpha\beta}^{ \mathbf{v}}\langle  c^{\mathbf{0}\dagger}_{\alpha\sigma}c_{\beta\sigma}^{\mathbf{v}}-c^{\mathbf{v}\dagger}_{\beta\sigma}c_{\alpha\sigma}^{\mathbf{0}}\rangle_{\phi},
    \label{eq:Jcurrent}
 \end{equation}
where the sites $(\alpha,\mathbf{0})$ and $(\beta,\mathbf{v})$ belong to different monolayers and the sum is restricted to one moir\'e cell.
The expectation value $\langle c^{\mathbf{0}\dagger}_{\alpha\sigma}c_{\beta\sigma}^{\mathbf{v}}-c^{\mathbf{v}\dagger}_{\beta\sigma}c_{\alpha\sigma}^{\mathbf{0}}\rangle_{\phi}$ is obtained by solving the gap equation self-consistently while imposing a fixed relative phase $\phi$ between the superconducting order parameters of the two monolayers. During this procedure, the gap amplitudes and chemical potential are allowed to converge self-consistently while the relative phase remains fixed. This approach provides a more accurate estimate of the Josephson current compared with our previous work \cite{panda2026}  and other  works, where the order parameters were not recalculated self-consistently after imposing the phase difference. The Josephson critical current is then defined as the maximum magnitude of the Josephson current over the phase difference,
\begin{equation}
    J_c=\max_{\phi}|J(\phi)|.
\end{equation}

We define the dimensionless quantity $j_c=(J_c\hbar a^2)/(eG_{\max}A)$ as the normalized Josephson critical current, where $A$ is the area of the moir\'e unit cell for a commensurate twist angle $\theta$. To remove the trivial scaling with the tunneling amplitude, we normalize by the maximum amplitude in the tunneling matrix, $G_{\max}$ (Fig.~\ref{g0vstheta}); for the phenomenological model, this would be $g_0$.
\par
\begin{figure}[tb]
\centering
\includegraphics[width=\linewidth]{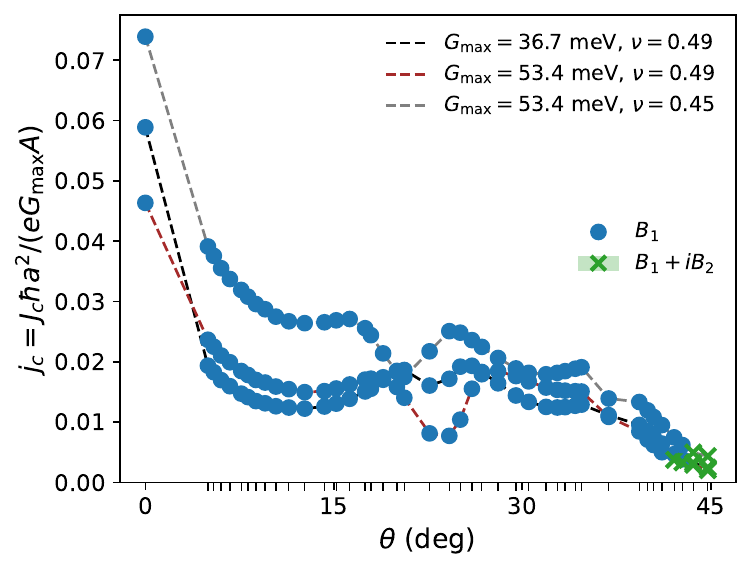}
\caption{
Josephson critical current $j_c=J_c\hbar a^2/(eG_{\max}A)$, calculated self-consistently, as a function of twist angle $\theta$ at fixed maximum interlayer tunneling amplitude $G_{\max}$. For each twist angle, the interlayer distance $c$ is adjusted to maintain the same $G_{\max}$ as that obtained at $43.6\degree$. The three parameter sets are $c=0.85a$ ($G_{\max}=36.7$~meV, $\nu=0.49$), represented by the black dashed line, and $c=0.75a$ ($G_{\max}=53.4$~meV, $\nu=0.49$ and $0.45$), represented by the red and gray dashed lines, respectively. The dashed lines connect the points corresponding to each parameter set, while the symbols indicate the superconducting ground-state symmetries. Additional commensurate twist angles are included to obtain a more detailed angle dependence of $j_c$. We find $B_1+iB_2$ states at all commensurate twist angles considered with $\theta\gtrsim42\degree$, accompanied by a strong suppression of the Josephson critical current.}
\label{jc_fixed_gmax_sc}
\end{figure}

The angle dependence of $j_c$ for three different parameter sets is shown in Fig.~\ref{jc_fixed_gmax_sc}. The three curves correspond to different combinations of filling $\nu$ and maximum tunneling amplitude $G_{\max}$. For each parameter set, we vary the interlayer distance $c$ to maintain a fixed $G_{\max}$ across different twist angles. To obtain a more detailed dependence of $j_c$ on $\theta$, we include additional commensurate twist angles beyond those considered in the phase diagram. At each twist angle, we first determine the interlayer distance $c$ corresponding to the desired $G_{\max}$ and then calculate the Josephson critical current self-consistently.

The cases shown are $G_{\max}=36.7$ meV and $\nu=0.49$, with the points connected by a black dashed line; $G_{\max}=53.4$ meV and $\nu=0.49$, with the points connected by a brown dashed line; and $G_{\max}=53.4$ meV and $\nu=0.45$, with the points connected by a gray dashed line. The symbols at each twist angle indicate the corresponding superconducting phases, as defined in the phase diagram in Fig.~\ref{Phase diagram}. We observe a sharp decrease in $j_c$ between $0\degree$ and $6\degree$, followed by a relatively weak angle dependence up to approximately $35\degree$, with an additional valley-type feature in the range $17\degree$--$26\degree$. As $\theta$ approaches $45\degree$, we find $B_1+iB_2$ states at all commensurate twist angles considered above approximately $42\degree$. In this regime, the critical current is suppressed further, reaching values approximately 15--20 times smaller than those at $0\degree$.  We have verified that the suppression of $j_c$ is primarily associated with changes in the Fermi surface rather than an overall reduction in the tunneling amplitudes or the superconducting order-parameter magnitudes.

\section{Discussion}

The advent of twistronics as a tuning knob for strongly correlated systems has provided a promising route toward realizing topological states with potential applications in quantum devices. Experimental techniques used to fabricate twisted cuprate junctions have advanced considerably, enabling the fabrication of clean samples exhibiting signatures consistent with TRSB chiral \mbox{$d+id'$} states. In order to make reliable predictions for future experimental probes, an accurate microscopic model that captures the essential physics of the cuprate monolayers is required. Ideally, \textit{ab initio} calculations that determine the Wannier wave functions for arbitrary commensurate twist angles while allowing for moir\'e-scale reconstruction would provide a more accurate microscopic description of these systems. Such calculations are beyond the scope of the present work but represent an important direction for future study. Most current models used to determine the ground-state properties of these systems, however, do not fully incorporate the relevant symmetries and orbital properties of the constituent monolayers. Wave-function modeling therefore provides an intermediate step toward understanding the changes that may arise as increasingly realistic microscopic descriptions of these systems are developed.
\par
 Another aspect of modeling these systems that requires careful investigation is the pairing interaction. Many theoretical models of twisted bilayer cuprates employ phenomenological nearest-neighbor attractive interactions to generate superconductivity. However, given the close connection between superconductivity and magnetic correlations in the cuprates, a more microscopic treatment of the origin and structure of the pairing interaction is necessary.
\par

\section{Conclusion}

In this work, we have taken a step toward incorporating the symmetry and sign-changing behavior of the BSCCO monolayer Wannier orbital by modeling the wave function at each Cu site using a $d_{x^2-y^2}$ wave function. We then calculated the orbital overlap between these localized wave functions and obtained an orbital-overlap-based tunneling model that differs significantly from the phenomenological model used in previous studies. We solved the self-consistent SC gap equation for two square lattices rotated by $\theta/2$ in opposite directions to form a given commensurate twist angle $\theta$. We evaluated the SC gap functions and projected them onto the irreducible representations of the $D_4$ point group to determine the corresponding order-parameter symmetries. We also compared the free energies of competing superconducting states to identify the energetically favored ground states in our phase diagram.
\par
\begin{figure*}[tb]
 \includegraphics[width=\linewidth]{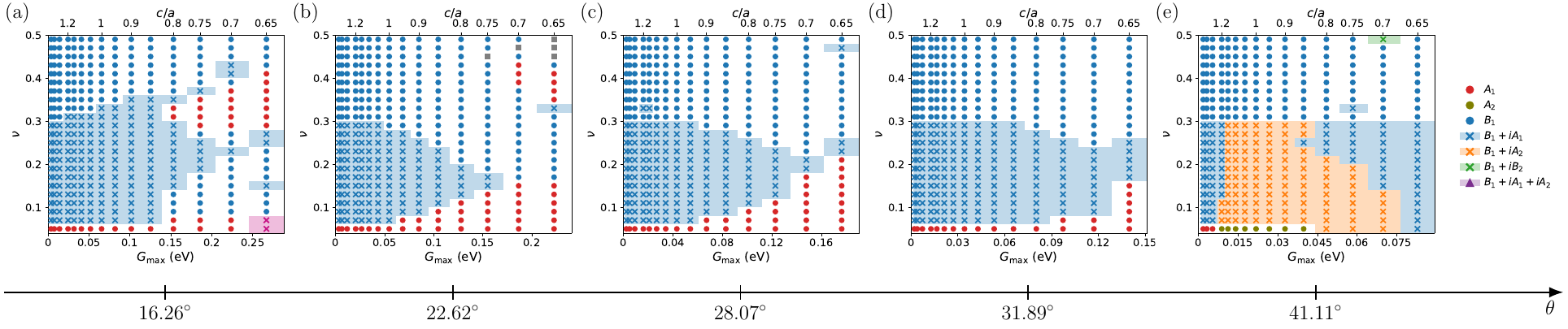}
    \caption{Phase diagrams of the orbital-overlap-based tunneling model for twist angles (a) $16.26\degree$, (b) $22.62\degree$, (c) $28.07\degree$, (d) $31.89\degree$, and (e) $41.11\degree$, as functions of the filling $\nu$ and the largest interlayer tunneling amplitude $G_{\max}$ (top axis: corresponding $c/a$). Parameters and symbols as in Fig.~\ref{Phase diagram}.}
    \label{Phase diagram other}
\end{figure*}
The phenomenological tunneling model can be represented by the overlap of isotropic $s$-wave wave functions, which produces tunneling amplitudes with the same sign that decay exponentially with the distance between lattice sites. In contrast, the orbital-overlap-based tunneling model exhibits distinctly different behavior, as demonstrated by the twist-angle dependence of the direct and maximum tunneling amplitudes in Fig.~\ref{g0vstheta} and the interlayer tunneling matrices $G_{\alpha\beta}^{\mathbf{v}}$ for $\theta=0\degree$ and $36.87\degree$ in Figs.~\ref{0_twist_tun} and \ref{hop-tun_53}, respectively. The orbital-overlap-based model produces sign-changing tunneling amplitudes, while the direct tunneling term remains the dominant amplitude at $0\degree$. As the twist angle approaches $45\degree$, the direct tunneling amplitude decreases due to the orbital mismatch between the wave functions in the two monolayers. Consequently, the off-vertical tunneling terms can dominate over the direct tunneling term, leading to considerable changes in the Fermi surfaces and the distribution of spectral weight. These changes, in turn, favor different SC ground states and result in substantially different phase diagrams as functions of filling, twist angle, and interlayer tunneling strength.
\par

In our previous work, we showed that the phenomenological phase diagram exhibited a \mbox{$d+id'$} state at a $28.07\degree$ twist for small direct tunneling amplitudes over a wide range of fillings. The region occupied by the \mbox{$d+id'$} phase grew as we considered commensurate twist angles closer to $45\degree$, consistent with previously calculated phase diagrams obtained from continuum models.

In contrast, in the orbital-overlap-based tunneling model presented here, the \mbox{$d+id'$} state is absent from most of the phase diagram for the commensurate twist angles considered up to $\theta=41.11\degree$. This indicates that the extent of the TRSB \mbox{$d+id'$} region predicted using the phenomenological tunneling model is substantially reduced when the orbital-dependent tunneling between the BSCCO layers is incorporated. Improvements in our computational methods have also allowed us to investigate substantially weaker tunneling amplitudes than were accessible in our previous work. These calculations suggest that a finite interlayer tunneling strength is necessary to stabilize a well-defined \mbox{$d+id'$} state. Together, these results highlight the importance of incorporating realistic orbital-dependent tunneling processes into microscopic models of twisted bilayer cuprates. Further improvements to the microscopic Hamiltonian may be essential for understanding the variability observed in experimental studies of twisted BSCCO junctions.
\par

\begin{acknowledgments}
A.K. acknowledges support by the Danish National Committee for Research Infrastructure (NUFI) through the ESS-Lighthouse Q-MAT. P.J.H. and S.K.P. were
supported by NSF-DMR-2231821. Computations were performed on resources in project UPPMAX 2026/1-3 provided by the National Academic Infrastructure for Supercomputing in Sweden (NAISS) at UPPMAX, funded by the Swedish Research Council through grant agreement no. 2022-06725.
\end{acknowledgments}

\appendix
\section{Phase diagrams at all commensurate twist angles}
\label{Phase_diagram_rest}
For reference, Fig.~\ref{Phase diagram other} shows the phase diagrams of the orbital-overlap-based tunneling model at the remaining commensurate twist angles not shown in Fig.~\ref{Phase diagram}. We see absence of the $B_1+iB_2$ states except a single point at $\theta=41.11\degree$.

\bibliography{references}
\end{document}